\documentclass[12pt]{article}
\usepackage{graphicx}
\usepackage{amsmath}
\usepackage{bm}       
\usepackage{latexsym} 
\usepackage{longtable}

\begin{document}

\baselineskip=14pt plus 1pt minus 1pt

{\Large \bf  Bohr Hamiltonian with deformation-dependent mass term}

\vskip 0.5truein

\begin{center}

Dennis Bonatsos$^1$, P. Georgoudis$^1$, D. Lenis$^1$, N. Minkov$^2$, and C. Quesne$^3$

\bigskip\bigskip

$^1$ Institute of Nuclear Physics, N.C.S.R. ``Demokritos'', \\ GR-15310 Aghia Paraskevi, Attiki, Greece

$^2$ Institute of Nuclear Research and Nuclear Energy, Bulgarian Academy of Sciences, 72 Tzarigrad Road, 1784 Sofia, Bulgaria

$^3$ Physique Nucl\'eaire Th\'eorique et Physique Math\'ematique, Universit\'e Libre de Bruxelles, Campus de la Plaine CP229,
Boulevard du Triomphe, B-1050 Brussels, Belgium

\vskip 0.5truein 

{\bf Abstract} 

\end{center} 

The Bohr Hamiltonian describing the collective motion of atomic nuclei is modified 
by allowing the mass to depend on the nuclear deformation. Exact analytical expressions 
are derived for spectra and wave functions in the case of a $\gamma$-unstable Davidson potential, 
using techniques of supersymmetric quantum mechanics. Numerical results in the Xe-Ba region are discussed.

\section{Introduction}

The Bohr Hamiltonian \cite{Bohr} and its extension, the geometrical 
collective model \cite{BM,EG}, have provided for several decades a sound framework 
for understanding the collective behaviour of atomic nuclei. A puzzle 
remaining unsolved since the early days of the Bohr Hamiltonian regards
the behaviour of the moments of inertia of atomic nuclei \cite{Ring}. 
They are predicted to increase proportionally to $\beta^2$, where $\beta$ is the collective 
variable corresponding to nuclear deformation, while experimentally 
a much more moderate increase is observed, especially for well deformed nuclei. 
In addition, the use of a constant mass has been recently questioned \cite{Jolos},
pointing out that the mass tensor of the collective Hamiltonian cannot be considered as
a constant and should be taken as a function of the collective coordinates. 

On the other hand, the algebraic framework of the Interacting Boson Model
(IBM) \cite{IA} has been very useful in the systematic study of nuclei corresponding
to its three limiting symmetries [vibrational U(5), axially symmetric 
deformed SU(3), $\gamma$-unstable O(6)], as well as to intermediate cases.
In the geometrical limit of the IBM \cite{IA}, obtained through the use 
of coherent states \cite{IA}, it is worth remembering that in addition 
to the usual term of the kinetic energy, $\pi^2$, terms of the form 
$\beta^2 \pi^2$ appear in the O(6) limiting symmetry and in the U(5)-O(6) transition region, 
while more complicated terms appear in the SU(3) limiting symmetry, as well as in the 
U(5)-SU(3) and SU(3)-O(6) transition regions \cite{vanRoos}. These terms indicate that it might
be appropriate to search for a modified form of the Bohr Hamiltonian, in
which the kinetic energy term will be modified by terms containing 
$\beta^2$, and even by more involved terms. 

The above reasoning leads to the consideration of a Bohr Hamiltonian with a 
mass depending on the collective variable $\beta$. 
Position-dependent effective masses have been considered recently 
in a general framework \cite{QT4267}, demonstrating the equivalence of this approach 
to the consideration of deformed canonical commutation relations, 
as well as to the consideration of curved spaces. Furthermore, several 
Hamiltonians known to be soluble through techniques of supersymmetric 
quantum mechanics (SUSYQM) \cite{PR,SUSYQM}, have been appropriately generalized \cite{Q2929} 
to include position-dependent effective masses, the 3-dimensional harmonic
oscillator being among them \cite{Q2929}.  

In the present work we are going to show that the Bohr Hamiltonian 
with a harmonic oscillator potential in $\beta$ 
(to which a term proportional to $1/\beta^2$ can be added at no cost)
can be generalized in order to include a mass depending on $\beta$,
$B=B_0/(1+a\beta^2)^2$, where $B_0$ and $a$ are constants. Exact solutions will
be constructed using techniques of SUSYQM. In order to achieve exact separation 
of variables, we are going to limit ourselves to potentials independent of the collective 
variable $\gamma$, called $\gamma$-unstable potentials. Numerical results for the 
Xe-Ba isotopes, well known examples of $\gamma$-unstable behaviour \cite{Casten}, will also be shown.


\section{Formalism of position-dependent effective masses} 

When the mass $m({\bf x})$ is position dependent \cite{QT4267}, 
it does not commute 
with the momentum ${\bf p} = -i\hbar \nabla$. As a consequence, there 
are many ways to generalize the usual form of the kinetic energy, 
${\bf p}^2 /(2 m_0)$, where $m_0$ is a constant mass, in order to obtain 
a Hermitian operator. In order to avoid any specific choices, one can use
the general two-parameter form proposed by von Roos \cite{vRoos},
with a Hamiltonian 
\begin{equation}
H = -{\hbar^2\over 4} [m^{\delta'}({\bf x}) \nabla m^{\kappa'}({\bf x})
\nabla m^{\lambda'}({\bf x}) + m^{\lambda'}({\bf x}) \nabla 
m^{\kappa'}({\bf x}) \nabla m^{\delta'}({\bf x})]+ V({\bf x}),
\end{equation}     
where $V$ is the relevant potential and the parameters $\delta'$, $\kappa'$,
$\lambda'$ are constrained by the condition $\delta'+ \kappa' + \lambda'=-1$.
Assuming  a position dependent mass of the  form 
\begin{equation}
m({\bf x}) = m_0 M({\bf x}), \qquad M({\bf x})=\frac{1}{(f({\bf x}))^2}, 
\qquad f({\bf x})=  1+ g ({\bf x}), 
\end{equation}
where $m_0$ is a constant mass and $M({\bf x})$ is a dimensionless 
position-dependent mass, the Hamiltonian becomes 
\begin{equation}
H = -{\hbar^2 \over 4 m_0} [f^{\delta}({\bf x}) \nabla f^{\kappa}({\bf x})
\nabla f^{\lambda}({\bf x})+ f^{\lambda}({\bf x}) \nabla f^{\kappa}({\bf x})
\nabla f^{\delta}({\bf x}) ] +V({\bf x}),
\end{equation}
with $\delta + \kappa +\lambda=2$. It is known \cite{QT4267} that this 
Hamiltonian can be put into the form 
\begin{equation}\label{eq:e3}
H =-{\hbar^2 \over 2 m_0} \sqrt{f({\bf x})} \nabla f({\bf x}) \nabla 
\sqrt{f({\bf x})} +V_{eff}({\bf x}), 
\end{equation} 
with
\begin{equation}
V_{eff}({\bf x}) = V({\bf x}) +{\hbar^2 \over  2 m_0} \left[ {1\over 2} 
(1-\delta-\lambda) f({\bf x}) \nabla^2 f({\bf x}) + \left({1\over 2}-\delta
\right) \left( {1\over 2}-\lambda\right) (\nabla f({\bf x}))^2 \right].  
\end{equation}

\section{Bohr Hamiltonian with deformation-dependent effective mass} 

The original Bohr Hamiltonian \cite{Bohr} is
\begin{equation}\label{eq:e1}
H = -{\hbar^2 \over 2B} \left[ {1\over \beta^4} {\partial \over \partial 
\beta} \beta^4 {\partial \over \partial \beta} + {1\over \beta^2 \sin 
3\gamma} {\partial \over \partial \gamma} \sin 3 \gamma {\partial \over 
\partial \gamma} - {1\over 4 \beta^2} \sum_{k=1,2,3} {Q_k^2 \over \sin^2 
\left(\gamma - {2\over 3} \pi k\right) } \right] +V(\beta,\gamma),
\end{equation}
where $\beta$ and $\gamma$ are the usual collective coordinates
($\beta$ being a deformation coordinate measuring departure from spherical shape, 
and $\gamma$ being an angle measuring departure from axial symmetry), while
$Q_k$ ($k=1$, 2, 3) are the components of angular momentum in the intrinsic 
frame, and $B$ is the mass parameter, which is usually considered constant.  

We wish to construct a Bohr equation with a mass depending on
the deformation coordinate $\beta$,
in accordance with the formalism described above,  
\begin{equation}
B(\beta)=\frac{B_0}{(f(\beta))^2}, 
\end{equation}
where $B_0$ is a constant. We then need the usual Pauli--Podolsky prescription
\cite{Podolsky} 
\begin{equation}
(\nabla \Phi)^i = g^{ij} {\partial \Phi \over \partial x^j}, \qquad 
\nabla^2 \Phi = {1\over \sqrt{g}} \partial_i \sqrt{g} g^{ij} \partial_j \Phi,
\end{equation}
in order to construct a Schr\"{o}dinger equation of the form of
Eq. (\ref{eq:e3}) 
in a 5-dimensional space equipped with the Bohr-Wheeler
coordinates $\beta,\gamma$. Since the deformation function
$f$ depends only on the radial coordinate $\beta$, 
only the $\beta$ part of the resulting equation will be 
affected, the final result reading
\begin{multline}\label{eq:mBohr}
\left[ 
-{1\over 2} {\sqrt{f}\over \beta^4} {\partial \over \partial \beta} 
\beta^4 f {\partial \over \partial \beta} \sqrt{f}
-{f^2 \over 2 \beta^2 \sin 3\gamma} {\partial \over \partial \gamma} 
\sin 3\gamma {\partial \over \partial \gamma} \right. \\
\left. + {f^2\over 8 \beta^2} 
\sum_{k=1,2,3} {Q_k^2 \over \sin^2\left(\gamma -{2\over 3} \pi k \right)}
+ v_{eff} \right] \Psi = \epsilon \Psi,  
\end{multline}
where reduced energies $\epsilon = B_0 E/\hbar^2$ and reduced potentials
$v= B_0 V/\hbar^2$ have been used, with
\begin{equation}
v_{eff}= v(\beta,\gamma)+ {1\over 4 } (1-\delta-\lambda) f \nabla^2 f 
+ {1\over 2} \left({1\over 2} -\delta\right) \left( {1\over 2} -\lambda\right)
(\nabla f)^2 .
\end{equation}


\section{The $\gamma$-unstable Davidson potential} 

The solution of the above Bohr-like equation can be reached for
certain classes of potentials using techniques developed in the
context of SUSYQM \cite{PR,SUSYQM,Q2929}. In order to achieve separation of variables
we assume that the potential $v(\beta,\gamma)$ depends only 
on the variable $\beta$, i.e. $v(\beta,\gamma)=u(\beta)$ \cite{Wilets}. 
Potentials of this kind are called $\gamma$-unstable potentials, since they are appropriate 
for the description of nuclei which can depart from axial symmetry without any energy cost. 
Furthermore, we are going to use as an example the Davidson potential \cite{Dav}
\begin{equation} \label{eq:e16}
u(\beta)=\beta^2 + {\beta_0^4\over \beta^2},
\end{equation}
where the parameter $\beta_0$ indicates the position of the minimum 
of the potential, the special case of $\beta_0=0$ corresponding to the 
simple harmonic oscillator. (Note that the term containing $\beta_0$ offers no additional complication 
to the solution). 

One then seeks wave functions of the form \cite{Wilets,IacE5}
\begin{equation}\label{fullwf}
\Psi (\beta, \gamma, \theta_i)= F(\beta) \Phi(\gamma, \theta_i),
\end{equation}
where $\theta_i$ ($i=1$, 2, 3) are the Euler angles.
Separation of variables gives
\begin{multline}\label{eq:e6}
\left[ 
-{1\over 2} {\sqrt{f}\over \beta^4} {\partial \over \partial \beta} 
\beta^4 f {\partial \over \partial \beta} \sqrt{f}
+{f^2\over 2\beta^2}  \Lambda 
+{1\over 4} (1-\delta-\lambda) f\nabla^2 f \right.\\ \left.  
+{1\over 2} \left( {1\over 2}-\delta\right) \left( {1\over 2} -\lambda\right)
(\nabla f)^2 +u(\beta) \right] F(\beta) = \epsilon F(\beta),  
\end{multline}
\begin{equation}\label{eq:e7}
\left[ -{1\over \sin 3\gamma} {\partial \over \partial \gamma} \sin 3\gamma
{\partial \over \partial \gamma} + {1\over 4} \sum_k {Q_k^2 \over 
\sin^2\left( \gamma -{2\over 3} \pi k\right) } \right] \Phi (\gamma, \theta_i)
= \Lambda \Phi(\gamma, \theta_i). 
\end{equation}
Eq. (\ref{eq:e7}) has been solved by B\`es \cite{Bes}. 
$ \Lambda=\tau(\tau+3)$
represents the eigenvalues of the second order Casimir operator of SO(5), 
while $\tau$ is the seniority quantum number, characterizing the irreducible
representations of SO(5). The values of angular momentum $L$ occurring 
for each $\tau$ are provided by a well known algorithm and are listed in \cite{IA,Wilets}.
Within the ground state band (gsb) one has $L=2\tau$. The $L=2$ member of the quasi-$\gamma_1$ band 
is degenerate with the $L=4$ member of the gsb, the $L=3$, 4 members of the quasi-$\gamma_1$ band 
are degenerate to the $L=6$ member of the gsb, the $L=5$, 6 members of the quasi-$\gamma_1$ band 
are degenerate to the $L=8$ member of the gsb, and so on.  

Eq. (\ref{eq:e6}) can be simplified by performing the derivations
\begin{equation}\label{eq:e10}
{1\over 2 } f^2 F''+ \left( f f'+{2 f^2\over \beta}\right) F'
+ \left( {(f')^2\over 8} + {f f''\over 4} +{f f'\over \beta}\right) F
-{f^2 \over 2\beta^2} \Lambda F
+\epsilon F -v_{eff}F=0, 
\end{equation}
with
\begin{equation}
v_{eff}= u + {1\over 4} (1-\delta-\lambda)  f \left( {4 f'\over \beta} 
+f''\right) + {1\over 2} \left( {1\over 2}-\delta\right) 
\left( {1\over 2}-\lambda\right) (f')^2. 
\end{equation}
The difference in the numerical coefficient of $f'$ observed in comparison 
to Eq. (2.27) of Ref. \cite{QT4267} is due to the different dimensionality 
of the space used in each case. 

Setting 
\begin{equation}\label{factorwf}
F(\beta)= {R(\beta)\over \beta^2}, 
\end{equation}
Eq. (\ref{eq:e10}) is put into the form 
\begin{equation}\label{eq:e14}
H R= -{1\over 2} \left(\sqrt{f}{d\over d\beta}\sqrt{f}\right)^2 R + u_{eff} R 
=\epsilon R, 
\end{equation}
where 
\begin{equation}\label{eq:e15}
u_{eff}= v_{eff} + {f^2+\beta f f'\over \beta^2}+ {f^2 \over 2 \beta^2}
\Lambda .
\end{equation}

Based on the results for the 3-dimensional harmonic oscillator reported 
in Ref. \cite{Q2929}, in order to find analytical results for Eq.
(\ref{eq:e14})
we are going to consider for the deformation function the special form 
\begin{equation}\label{eq:e17}
f(\beta)=1+a \beta^2.
\end{equation}
This choice is made in order to lead to an exact solution. Its physical 
implications will be discussed in Section 8.

Using these forms for the potential and the deformation function 
one obtains 
\begin{multline}\label{eq:e14b}
u_{eff}= \beta^2  +a^2\beta^2\left[  {5\over 2} (1-\delta-\lambda)  
+ 2 \left({1\over 2}-\delta\right) \left({1\over 2}-\lambda\right)  
+ 3  + {\Lambda\over 2}\right] \\
+ {1\over \beta^2} \left( 1+{1\over 2} \Lambda +\beta_0^4\right) 
+a\left[ {5\over 2} (1-\delta -\lambda)  + 4+  \Lambda  \right]. 
\end{multline}

\section{Factorization}\label{sec:fact}

Following the general method used in supersymmetric quantum mechanics (SUSYQM) \cite{PR,SUSYQM}, 
one should take the following steps:

i) Factorize the Hamiltonian. 

ii) Write a hierarchy of Hamiltonians starting from the first one.

iii) Impose the shape invariance conditions, which are integrability conditions 
guaranteeing exact solvability. 

Thus one first tries to put the Hamiltonian in Eq. (\ref{eq:e14}) in the form 
\begin{equation}\label{eq:e17b}
H_0=B_0^+ B_0^- +\varepsilon_0.  
\end{equation} 
Then this Hamiltonian can be considered as the first member of a hierarchy 
of Hamiltonians
\begin{equation}\label{eq:e18}
H_i = B_i^+ B_i^- + \sum _{j=0}^i \varepsilon_j, 
\end{equation} 
expressed in terms of the generalized ladder operators $B_i^+$, $B_i^-$,
which are  determined recursively, along with $\varepsilon_i$ 
[which are in general different from the energy eigenvalues $\epsilon$ appearing in Eq. (\ref{eq:e14}),
the only exception being $\varepsilon_0=\epsilon_0$], 
through the shape invariance (SI) condition 
\begin{equation}\label{eq:e19}
B_i^- B_i^+= B_{i+1}^+ B_{i+1}^- +\varepsilon_{i+1} . 
\end{equation}
In the  present case one can start with
\begin{equation}\label{eq:e18b}
B_0^{\pm}= \mp{1\over \sqrt{2}} \left(\sqrt{f}{d\over d\beta} \sqrt{f}\right) 
+{1\over \sqrt{2}} \left(c_0 \beta + \bar c_0 {1\over \beta}\right) , 
\end{equation} 
where the second term resembles the superpotential occurring in the 
case of the 3-dimensional harmonic oscillator, reported in Ref. 
\cite{Q2929}. 
Substituting into Eq. (\ref{eq:e17b}) one obtains 
\begin{equation}\label{eq:e20} 
H_0 = -{1\over 2} \left( \sqrt{f} {d \over d\beta} \sqrt{f} \right)^2 
+ {\beta^2 \over 2} (c_0^2-a c_0) + {(\bar c_0^2 + \bar c_0)\over 2\beta^2} 
 + \left(-{1\over 2} c_0 + {1\over 2} a \bar c_0
+ c_0 \bar c_0 +\varepsilon_0\right) .
\end{equation} 
Comparing this result to Eq. (\ref{eq:e14}) with the effective potential 
of Eq. (\ref{eq:e14b}) and equating powers of $\beta$ one gets the following:
\begin{equation}\label{eq:e21}
c_0= {1\over 2} (a \pm \sqrt{a^2+8 P_1}) ,
\end{equation}
\begin{equation}\label{eq:e22}
\bar c_0  = {1\over 2} (-1\pm \sqrt{9+ 4\Lambda+ 8 \beta_0^4}),\quad  \text{and}
\end{equation}
\begin{equation}\label{eq:e22a}
\varepsilon_0= {1\over 2} c_0 - {1\over 2} a \bar c_0 - c_0 \bar c_0 
+{5\over 2} (1-\delta-\lambda) a + 4a + a\Lambda,
\end{equation}
where
\begin{equation}
P_1= 1+ a^2 \left[ {5\over 2} (1-\delta-\lambda) + 2\left({1\over 2}-\delta
\right) \left( {1\over 2}-\lambda\right) + 3 + {\Lambda\over 2}  \right].
\end{equation}

Upon substitution of $c_0$ and $\bar c_0$ from Eqs. (\ref{eq:e21}) and 
(\ref{eq:e22}), Eq. (\ref{eq:e22a}) leads to
\begin{multline}\label{Egs}
\varepsilon_0 = a\left(\frac{29}{4} -{5\over 2}(\delta+\lambda) +\Lambda\right) 
\pm {1\over 2} \sqrt{a^2+8 P_1} \mp {a\over 2} \sqrt{9+4\Lambda+8 \beta_0^4} \\
+\sigma {1\over 4} \sqrt{(a^2+8 P_1)(9 + 4\Lambda+ 8 \beta_0^4)},
\end{multline}
where $\pm$ refers to the sign in Eq. (\ref{eq:e21}), $\mp$ corresponds 
to the sign selection in Eq. (\ref{eq:e22}), while $\sigma$ is $+1$ 
for the selections $(+,-)$ and $(-,+)$ in Eqs. (\ref{eq:e21}) and 
(\ref{eq:e22}) respectively, while it is $-1$ for the selections 
$(+,+)$ and $(-,-)$. 

The limiting case of $a=0$, in which $P_1=1$, can be used for reducing the number 
of possibilities. Since $\varepsilon_0$ should be increasing with $\Lambda$ 
(members of the ground state band should exhibit energies increasing with the angular momentum $L$), 
it turns out that one should have $\sigma=+1$,
thus only the choices $(+,-)$ and $(-,+)$ are acceptable.   

In Section \ref{wf}  it will be shown that good behaviour of the wavefunctions 
at $\beta=0$ leaves $(+,-)$ as the only choice.

\section{Shape invariance}\label{shape}

In the next step the hierarchy of Hamiltonians of Eq. (\ref{eq:e18})
should be considered, with
\begin{equation}
B_i^{\pm}= \mp {1\over \sqrt{2}}\left( \sqrt{f} {d\over d\beta} \sqrt{f}
\right)+ {1\over \sqrt{2}} \left(c_i \beta + {\bar c_i \over \beta}\right). 
\end{equation} 
Substituting these expressions in the shape invariance condition of 
Eq. (\ref{eq:e19}) and equating powers of $\beta$
leads to the following results:
\begin{equation}\label{eq:ci}
c_i^2 + c_ia = c_{i+1}^2 -c_{i+1}a ,
\end{equation}
\begin{equation}
\bar c_i^2 -\bar c_i = \bar c_{i+1}^2 +\bar c_{i+1},
\end{equation}
and
\begin{equation}
2\varepsilon_{i+1}= c_i + c_{i+1} -a \bar c_i -a \bar c_{i+1} + 2 c_i \bar c_i 
-2  c_{i+1} \bar c_{i+1}. 
\end{equation}
Keeping from the first two of these only the solutions 
 $c_{i+1}=c_i+a$ (leading to $c_i = c_0+i a$) and  $\bar c_{i+1}= \bar c_i-1$
 (leading to $\bar c_i = \bar c_0-i$),  
in accordance 
with the results obtained for the 3-dimensional harmonic oscillator
\cite{Q2929}, we get 
\begin{equation}
\varepsilon_i = 2 [c_0 -a\bar c_0 +a(2i-1)]. 
\end{equation} 
One then easily finds for the energy  
\begin{equation}\label{Enu}
\epsilon_\nu = \sum_{i=0}^\nu \varepsilon_i =\epsilon_0 + 2\nu(c_0-a\bar c_0)
+ 2 a \nu^2.
\end{equation}
The ground state band is obtained for $\nu=0$, while the quasi-$\beta_1$ band 
corresponds to $\nu=1$. 

Equation (\ref{Enu}) only provides a formal solution to the bound-state energy spectrum. 
The range of $\nu$ values is actually determined by the existence of corresponding physically acceptable wavefunctions,
to be discussed in the next Section. 

\section{Wave functions}\label{wf}

To be physically acceptable, the bound-state wavefunctions should satisfy two conditions \cite{Q2929}:

\noindent (i) As in conventional (constant-mass) quantum mechanics, they should be square integrable on
the interval of definition of $u_{\rm eff}$, i.e.,
\begin{equation}
  \int_{0}^{\infty} d\beta\,  |R_\nu(\beta)|^2 < \infty.  \label{eq:wf-C1}
\end{equation}
\noindent (ii) Furthermore, they should ensure the Hermiticity of $H$. For such a purpose, it is enough to impose that the operator $\sqrt{f} (d/d\beta) \sqrt{f}$ be Hermitian, which amounts to the restriction
\begin{equation}
  |R_\nu(\beta)|^2 f(\beta) \to 0 \qquad {\rm for\ } \beta \to 0 {\rm \ and\ } \beta \to \infty,
\end{equation}
or, equivalently,
\begin{equation}
  |R_\nu(\beta)|^2 \to 0 \quad {\rm for\ } \beta \to 0 \qquad {\rm and} \qquad |R_\nu(\beta)|^2 \beta^2 
  \to 0 \quad {\rm for\ } \beta \to \infty.  \label{eq:wf-C2}
\end{equation}
As condition (\ref{eq:wf-C2}) is more stringent than condition (\ref{eq:wf-C1}), we should only be concerned with the former.

The ground state wave function can be determined from the differential 
equation \cite{Q2929} 
\begin{equation}
B^-_0(\beta; c_0, \bar c_0) R_0 (\beta; c_0, \bar c_0) = 0, 
\end{equation}
where $B^-_0$ is given by Eq. (\ref{eq:e18b}). 
Trying the solution 
\begin{equation}\label{gsbwf}
R_0  = C_0 \beta^n f^{\bar n},
\end{equation}
where $C_0$ is a normalization constant, 
the powers of $\beta$ lead to the conditions 
\begin{equation}
n=-\bar c_0,
\end{equation}
and
\begin{equation}
\bar n = {a \bar c_0 - c_0 -a\over 2 a}. 
\end{equation}

For $\beta \to 0$, the function $|R_0(\beta)|^2$ behaves as $\beta^{2n}$.
Condition (\ref{eq:wf-C2}) imposes that $n > 0$, i.e. $\bar c_0 <0$. 
From Eq. (\ref{eq:e22}) it is clear that this is 
guaranteed if the minus sign is retained in it. From the discussion 
given at the end of Section \ref{sec:fact} it is clear that the only possibility 
remaining for the signs in Eqs. (\ref{eq:e21}) and (\ref{eq:e22}) 
is the $(+,-)$ one. One can easily see that this choice guarantees
$\bar n<0$. 

For $\beta \to \infty$, $|R_0(\beta)|^2 \beta^2$ behaves as $\beta^{- 2c_0/a}$, since in this case $f\approx a \beta^2$.
Condition (\ref{eq:wf-C2}) therefore imposes that $c_0 > 0$. This restriction is already satisfied,
since we have been led in the previous paragraph to keep the upper sign choice in (\ref{eq:e21}). 

Wave functions of excited states can then be obtained from the 
recursion relation \cite{Q2929} 
\begin{equation}
R_{\nu+1}(\beta; c_0, \bar c_0) = 
{1\over \sqrt{\epsilon_{\nu+1}(c_0, \bar c_0)-\epsilon_0(c_0, \bar c_0)}} 
B^+(\beta; c_0, \bar c_0)  R_\nu(\beta; c_1, \bar c_1),   
\end{equation}
where the different coefficients appearing in the last term should 
be noticed. From the recursion relation one obtains 
\begin{equation}\label{b1wf}
R_1(\beta; c_0, \bar c_0)= {C_1 \over 2^{1/2} \sqrt{\epsilon_1-\epsilon_0}}
\left[ (2 c_0+a)\beta +{2 \bar c_0-1\over \beta}\right] \beta^{n+1} 
f^{\bar n-1}, 
\end{equation}
\begin{multline}
R_2(\beta; c_0, \bar c_0)= {C_2 \over 2^{3/2} \sqrt{(\epsilon_2-\epsilon_0)(\epsilon_1-\epsilon_0)}} \\
\left[ (2 c_0+ 3 a)(2 c_0 + a) \beta^2 + (2 \bar c_0 -3) (2 \bar c_0-1)
{1\over \beta^2} + 2(2 c_0+3a)(2\bar c_0-3)\right] \beta^{n+2} f^{\bar n-2},
\end{multline}
where $C_1$, $C_2$ are normalization constants. 

From the above it is clear that wave functions of the states belonging to the ground state band
are obtained from Eq. (\ref{fullwf}), substituting in it Eqs. (\ref{factorwf}) and (\ref{gsbwf}),  
while wave functions of the states belonging to the quasi-$\beta_1$ band
are obtained from Eq. (\ref{fullwf}), substituting in it Eqs. (\ref{factorwf}) and (\ref{b1wf}).   

It is easy to see that the conditions imposed above in order to guarantee the physically acceptable 
behaviour of the ground state wavefunctions, also guarantee the physically acceptable behaviour 
of the wavefunctions for excited states. In particular:

\noindent (i) In order to examine the behaviour at $\beta\to 0$, it suffices to examine in $R_{\nu}$ the behaviour 
of the polynomial term containing the lowest power of $\beta$ ($\beta^{-1}$ in $R_1$, $\beta^{-2}$ in $R_2$). 
In both cases the function $|R_\nu(\beta)|^2$ behaves as $\beta^{2n}$, 
i.e. it exhibits the same behaviour as $|R_0(\beta)|^2$.

\noindent (ii) In order to examine the behaviour at $\beta\to \infty$, it suffices to examine in $R_{\nu}$ the behaviour 
of the polynomial term containing the highest power of $\beta$ ($\beta$ in $R_1$, $\beta^2$ in $R_2$). 
In both cases the function $|R_\nu(\beta)|^2 \beta^2 $ behaves as $\beta^{- 2c_0/a}$, 
i.e. it exhibits the same behaviour as $|R_0(\beta)|^2 \beta^2$.

Therefore the wavefunctions of the excited states given above are forced to exhibit physically acceptable behaviour
by the same conditions which guarantee the physically acceptable behaviour of the ground state wavefunctions.   

\section{Numerical results}

From Eq. (\ref{eq:mBohr}) it is clear that in the present case the moments of inertia 
are not proportional to $\beta^2 \sin^2\left( \gamma -2 \pi k/3\right)$ but to 
$(\beta^2/f^2(\beta)) \sin^2\left( \gamma -2 \pi k/3\right)$. The function $\beta^2/ 
f^2(\beta)$ is shown in Fig. 1 for different values of the parameter $a$. It is clear 
that the increase of the moment of inertia is slowed down by the function $f(\beta)$,
as it is expected as nuclear deformation sets in \cite{Ring}.  

As a first testground of the present method we have used the Xe isotopes shown in Table 1. 
They have been chosen because: 

i) They are known to lie in a $\gamma$-unstable region \cite{Casten}.  

ii) At least the bandheads of the quasi-$\beta_1$ and quasi-$\gamma_1$ bands are known experimentally.

iii) They extend from the borders of the neutron shell ($^{134}$Xe$_{80}$ is just below the N=82 shell closure)
to the midshell ($^{120}$Xe$_{66}$) and even beyond, exhibiting increasing collectivity (increasing $R_{4/2}=E(4_1^+)/E(2_1^+)$ ratios)
from the border to the mishell. 

For evaluating the rms fits performed, the quality measure 
\begin{equation}\label{eq:e99}
\sigma = \sqrt{ { \sum_{i=1}^n (E_i(exp)-E_i(th))^2 \over
(n-1)E(2_1^+)^2 } }
\end{equation}
has been used. 
The theoretical predictions for the levels of the ground state band are obtained from Eq. (\ref{Egs})
(in which all terms with double signs are taken with positive signs, as explained at the end of Section 5), 
while the levels of the quasi-$\beta_1$ band are obtained from Eq. (\ref{Enu}) for $\nu=1$.
The levels of the quasi-$\gamma_1$ band are obtained through their degeneracies to members of the ground state band,
mentioned below Eq. (\ref{eq:e7}). 

Moving from the border of the neutron shell to the midshell, the following remarks apply

i) $^{134}$Xe and $^{132}$Xe are almost pure vibrators. Therefore no need for deformation dependence of the mass
exists, the least square fitting leading to $a=0$. Furthermore, no $\beta_0$ term is needed in the potential, 
the fitting therefore leading to $\beta_0=0$, {\it i.e.}, to pure harmonic behaviour. 

ii) In the next two isotopes ($^{130}$Xe and $^{128}$Xe) the need to depart from the pure harmonic oscillator 
becomes clear, the fitting leading therefore to nonzero $\beta_0$ values. However, there is still no need 
of dependence of the mass on the deformation, the fitting still leading to $a=0$. 

iii) Beyond $^{126}$Xe both the $\beta_0$ term in the potential and the deformation dependence of the mass become 
necessary, leading to nonzero values of both $\beta_0$ and $a$. 

Exactly the same behaviour is seen in the Ba isotopes, also known to lie in a $\gamma$-unstable region \cite{Casten}
and shown in Table 1. 

The results shown in Table 1 have been obtained for $\delta=\lambda=0$. One can easily verify that 
different choices for $\delta$ and $\lambda$ lead to a renormalization of the parameter values $a$ and $\beta_0$, 
the predicted energy levels remaining practically the same. 

In addition to energy spectra, B(E2) transition rates should be calculated and compared to experiment. 
The details of this task are deferred to a longer publication. However, basic qualitative features 
can be seen in Fig.~2, where the systematic behaviour of energy ratios and B(E2) ratios within the ground state band are shown. 
We remark that an increase of the $a$ parameter leads to a more rapid increase of energies (normalized to the energy of the first 
excited state) within the ground state band as a function of the angular momentum $L$, while in parallel it slows down 
the increase of B(E2)s (normalized to the transition from the first excited state to the ground state) as a function of $L$. 

Among the nuclei of Table 1, the only one exhibiting experimentally known increasing B(E2)s within the ground state band 
is $^{128}$Xe. The theoretical predictions (using the parameters of Table 1, obtained by fitting the energy levels alone)
fall withing the experimental error bars, as seen in Table 2.

The present results suggest that dependence of the mass on deformation becomes necessary 
as deformation increases. It is therefore desirable to provide a similar solution of the Bohr Hamiltonian 
applicable to axially symmetric well deformed nuclei. Work in this direction is in progress.

\section{Conclusion}

Motivated by the existence in the geometrical limit of the O(6) limiting symmetry and of the U(5)-O(6) transition region 
of the Interacting Boson Model of extra terms 
of the form $\beta ^2 \pi^2$ (where $\beta$ is the nuclear deformation) in addition to 
the kinetic energy term $\pi^2$, as well as by the existence of more complicated additional terms 
in the SU(3) limiting symmetry and in the U(5)-SU(3) and SU(3)-O(6) transition regions, 
we have modified the Bohr Hamiltonian describing the collective 
motion of atomic nuclei by allowing the mass to depend on the nuclear deformation. 
Using techniques of supersymmetric quantum mechanics we have obtained exact analytical expressions 
for spectra and wave functions for the case of the $\gamma$-unstable Davidson potential.
A first numerical application in the Xe-Ba region gives encouraging results.  
Detailed comparisons to experiment, including B(E2) transition rates, as well as extension of the method to 
deformed axial nuclei (with $\gamma \approx 0$) are deferred to a longer publication. 

{\bf Acknowledgements}

The authors are thankful to F. Iachello for suggesting the project and for useful discussions.
One of the authors (N. M.) acknowledges the support of the Bulgarian Scientific Fund under contract F-1502/05.

\begin{table}

\caption{Comparison of theoretical predictions of the 
$\gamma$-unstable Bohr Hamiltonian with $\beta$-dependent mass (with $\delta =\lambda =0$)
to experimental data \cite{NDS} of Xe and Ba isotopes.  
The $R_{4/2}=E(4_1^+)/E(2_1^+)$ ratios, as well as the quasi-$\beta_1$ and quasi-$\gamma_1$ bandheads, normalized
to the $2_1^+$ state and labelled by $R_{0/2}=E(0_{\beta}^+)/E(2_1^+)$ and $R_{2/2}=E(2_{\gamma}^+)/E(2_1^+)$ respectively, are shown. 
The angular momenta of the highest levels of the ground state, quasi-$\beta_1$ and quasi-$\gamma_1$ bands included in the rms fit are labelled by $L_g$,
$L_\beta$, and $L_\gamma$ respectively, while $n$ indicates the total number of levels involved in the fit and $\sigma$ is the
quality measure of Eq. (\ref{eq:e99}). The theoretical predictions are obtained from the formulae mentioned below Eq. (\ref{eq:e99}).
See Section 8 for further discussion. }

\bigskip

\begin{tabular}{ r r r r  r r r r  r r r r r r}
\hline nucleus & $R_{4/2}$ & $R_{4/2}$ & $R_{0/2}$& $R_{0/2}$ & $R_{2/2}$ & $R_{2/2}$ & $\beta_0$ & $a$ & 
$L_g$ & $L_\beta$ & $L_\gamma$ & $n$ & $\sigma$ \\
        & exp &  th  & exp & th  & exp & th &  &  &  &  &  &   &  \\

\hline

$^{118}$Xe & 2.40 & 2.32 & 2.5 & 2.6 & 2.8 & 2.3 & 1.27 & 0.103 & 16 & 4 &10 & 19 & 0.319 \\
$^{120}$Xe & 2.47 & 2.36 & 2.8 & 3.4 & 2.7 & 2.4 & 1.51 & 0.063 & 26 & 4 & 9 & 23 & 0.524 \\
$^{122}$Xe & 2.50 & 2.40 & 3.5 & 3.3 & 2.5 & 2.4 & 1.57 & 0.096 & 16 & 0 & 9 & 16 & 0.638 \\
$^{124}$Xe & 2.48 & 2.36 & 3.6 & 3.5 & 2.4 & 2.4 & 1.55 & 0.051 & 20 & 2 &11 & 21 & 0.554 \\
$^{126}$Xe & 2.42 & 2.33 & 3.4 & 3.1 & 2.3 & 2.3 & 1.42 & 0.064 & 12 & 4 & 9 & 16 & 0.584 \\
$^{128}$Xe & 2.33 & 2.27 & 3.6 & 3.5 & 2.2 & 2.3 & 1.42 & 0.000 & 10 & 2 & 7 & 12 & 0.431 \\
$^{130}$Xe & 2.25 & 2.21 & 3.3 & 3.1 & 2.1 & 2.2 & 1.27 & 0.000 & 14 & 0 & 5 & 11 & 0.347 \\
$^{132}$Xe & 2.16 & 2.00 & 2.8 & 2.0 & 1.9 & 2.0 & 0.00 & 0.000 &  6 & 0 & 5 &  7 & 0.467 \\
$^{134}$Xe & 2.04 & 2.00 & 1.9 & 2.0 & 1.9 & 2.0 & 0.00 & 0.000 &  6 & 0 & 5 &  7 & 0.685 \\
           &      &      &     &     &     &     &      &       &    &   &   &    &       \\
$^{130}$Ba & 2.52 & 2.42 & 3.3 & 3.2 & 2.5 & 2.4 & 1.60 & 0.118 & 12 & 0 & 6 & 11 & 0.352 \\  
$^{132}$Ba & 2.43 & 2.29 & 3.2 & 2.8 & 2.2 & 2.3 & 1.29 & 0.059 & 14 & 0 & 8 & 14 & 0.619 \\
$^{134}$Ba & 2.32 & 2.16 & 2.9 & 2.7 & 1.9 & 2.2 & 1.12 & 0.000 &  8 & 0 & 4 &  7 & 0.332 \\
$^{136}$Ba & 2.28 & 2.00 & 1.9 & 2.0 & 1.9 & 2.0 & 0.00 & 0.000 &  6 & 0 & 2 &  4 & 0.250 \\  

\hline
\end{tabular}
\end{table}

\begin{table}

\caption{Experimental $B(E2; L\to L-2)$ transition rates [normalized to the transition from the first excited state to the ground state, 
$B(E2;2\to 0)$] within the ground state band of $^{128}$Xe \cite{NDS}, compared to theoretical predictions using the parameters 
of Table 1. See Section 8 for further discussion.} 

\bigskip

\begin{tabular}{ r r r }
\hline 
$L$ & exp. & th. \\
\hline
4  & $1.468\pm 0.201$ & 1.632 \\ 
6  & $1.940\pm 0.275$ & 2.196 \\
8  & $2.388\pm 0.398$ & 2.751 \\
10 & $2.736\pm 1.138$ & 3.310 \\
\hline
\end{tabular}
\end{table}


\begin{figure}[ht]
\centering
\includegraphics[height=6cm ]{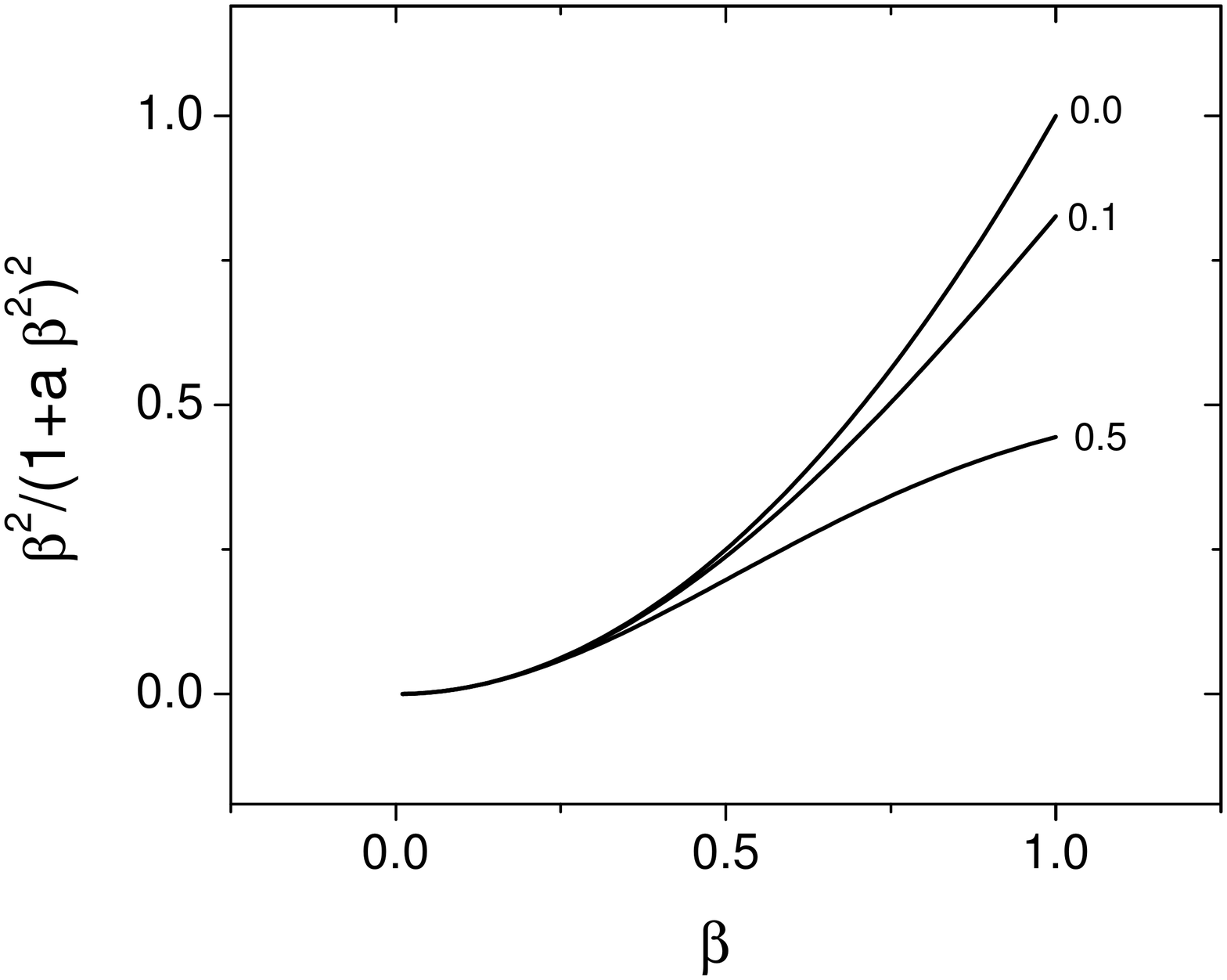}
\caption{The function $\beta^2 / f^2(\beta) = \beta^2 / (1+a \beta^2)^2$, to which 
moments of inertia are proportional as seen from Eq. (\ref{eq:mBohr}), plotted 
as a function of the nuclear deformation $\beta$ for different values of the parameter $a$.
See Section 8 for further discussion.}
\end{figure}


\begin{figure}[ht]
\centering
\includegraphics[height=6cm ]{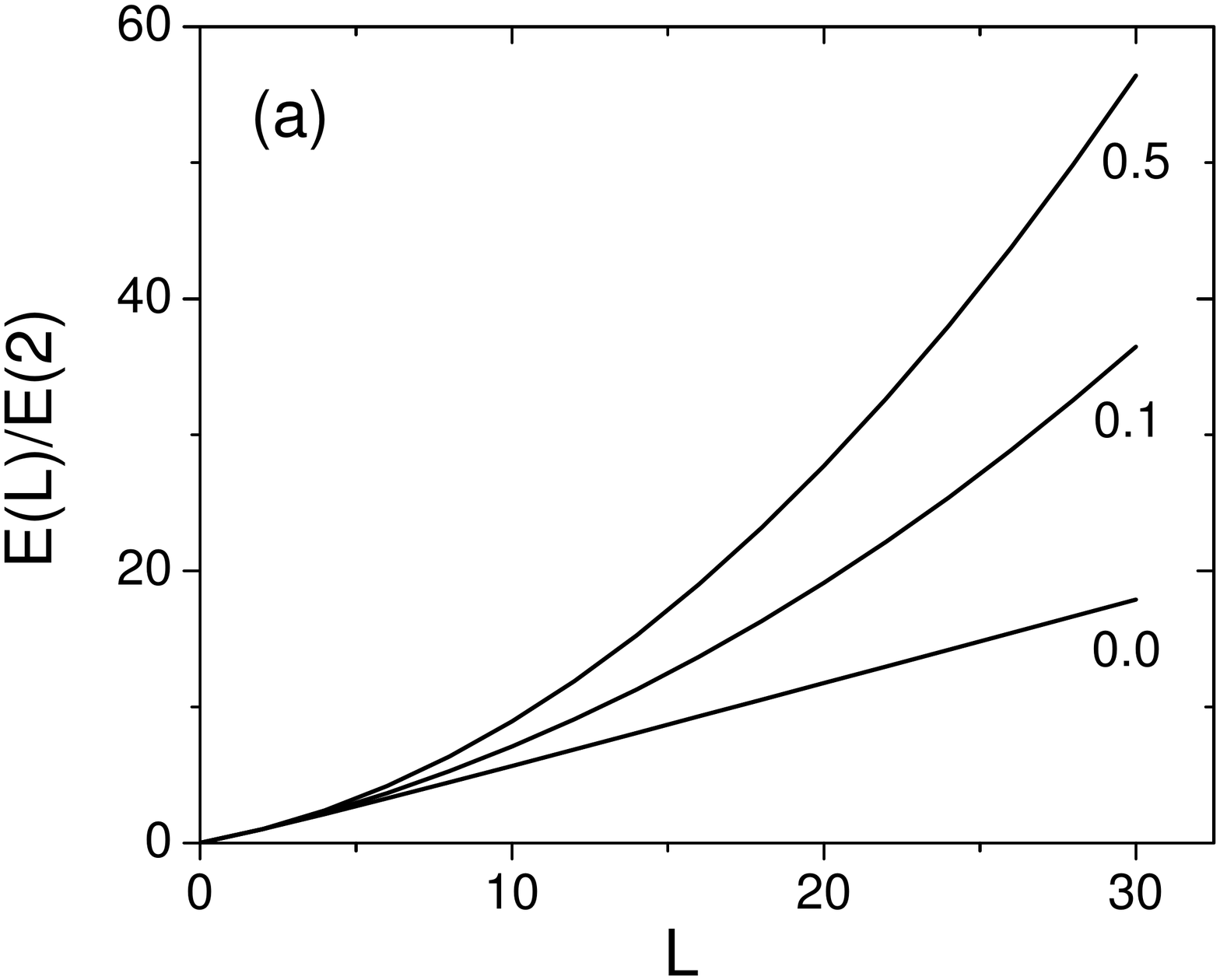}
\includegraphics[height=6cm ]{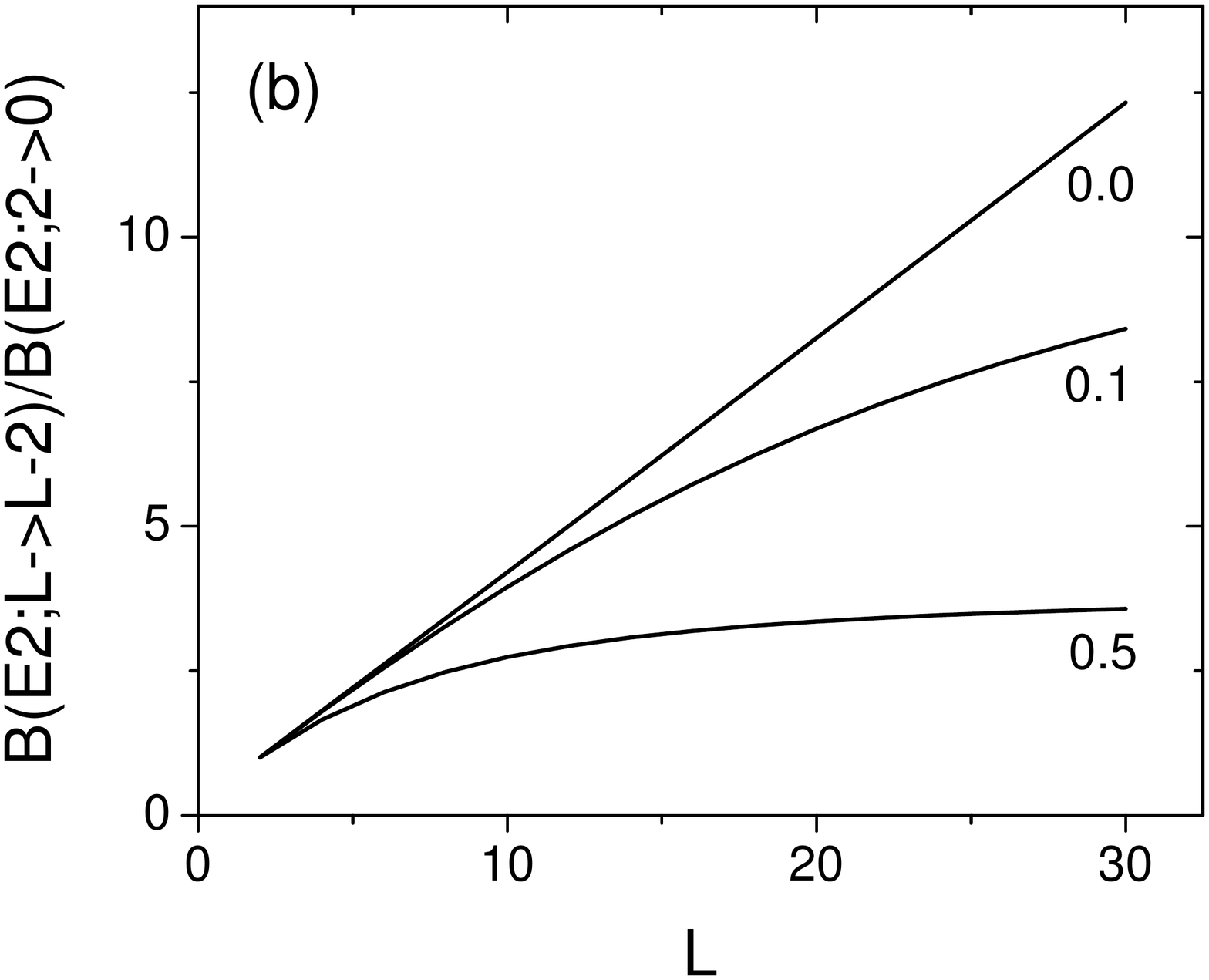}
\caption{Energy levels $E(L)$ [normalized to the energy of the first excited state, E(2)]
and $B(E2; L\to L-2)$ transition rates [normalized to the transition from the first excited state to the ground state, 
$B(E2;2\to 0)$] are shown for the ground state band as functions of the angular momentum $L$ for $\beta_0=1$
and varying values of $a$ (0.0, 0.1, 0.5).  See Section 8 for further discussion.} 

\end{figure}

\end{document}